\documentclass[aps,eqsecnum,nofootinbib,superscriptaddress,showpacs]{revtex4}
\usepackage[dvips]{graphicx}  % for graphics
\usepackage{amsmath,amsfonts}
\usepackage{bm}
\usepackage{color}  % for color
\def\be{\begin{eqnarray}}
\def\ee{\end{eqnarray}}
\def\beq{\begin{equation}}
\def\eeq{\end{equation}}

\def\p{\partial}

\def\({\left (}
\def\){\right )}
\def\[{\left [}
\def\[{\right ]}

\newcommand{\Exp}[1]{\langle\, #1\, \rangle}

\bmdefine{\bmk}{\bm{k}}
\bmdefine{\bmx}{\bm{x}}
\newcommand{\bbS}{\mathbb{S}}
\begin{document}
\title{Radiation from an accelerated quark via AdS/CFT} 
\author{Kengo Maeda}
\email{maeda302@sic.shibaura-it.ac.jp}
\affiliation{Department of Engineering, 
Shibaura Institute of Technology, Saitama, 330-8570, Japan}

%\author{Makoto Natsuume}
%\email{makoto.natsuume@kek.jp}
%\affiliation{Theory Division, Institute of Particle
%and Nuclear Studies KEK, \\
%High Energy Accerelator Research Organization,
%Tsukuba, Ibaraki, 305-0801, Japan}
\author{Takashi Okamura}
\email{tokamura@kwansei.ac.jp}
\affiliation{Department of Physics, Kwansei Gakuin University,
Sanda, 669-1337, Japan}

\date{\today}
\begin{abstract}
In this paper we investigate radiation by an accelerated quark
in a strongly coupled gauge theory via AdS/CFT correspondence. 
According to AdS/CFT dictionary, we can read off energy density
or energy flux of the radiation
from asymptotic gravitational field in AdS bulk
sourced by an accelerated string trailing behind the quark.
In the case of an oscillating quark with frequency $\Omega$,
we show that the time averaged energy density
is asymptotically isotropic and it falls off as
%$\sqrt{g_{\text{YM}}^2 N}\, \Omega^4/R^{2}$ with distance $R$
$(g_{\text{YM}}^2 N)^{1/2}\, \Omega^4/R^{2}$ with distance $R$
from the source.
In a toy model of a scattered quark by an external field,
we simply estimate Poynting vector by the bremsstrahlung radiation
and show that the energy flux is anisotropic outgoing radiation.
Based on these investigations,
we discuss the properties of strongly coupled gauge theory radiation
in comparison with electromagnetic radiation.
%Both models suggest that the bremsstrahlung radiation in a strongly
%coupled gauge theory is very similar to
%the bremsstrahlung radiation of photon in electromagnatic
%field except the angular distribution of the radiation.
\end{abstract}
\maketitle

%%%%%%%%%%%%%%%%%%%%%%%%%%%%%%%%%%%%%%%%%%%%%%%%%%%%%%%%%%%%%
\section{Introduction}\label{sec:intro}
%%%%%%%%%%%%%%%%%%%%%%%%%%%%%%%%%%%%%%%%%%%%%%%%%%%%%%%%%%%%%
It is quite interesting to investigate
the mechanism of energy dissipation by an external quark moving
through gauge theory plasma from both theoretical
and experimental point of view.
Recently, a strongly coupled gauge theory plasma
is believed to be formed at RHIC~\cite{Shuryak04,Shuryak05}.
A heavy quark moving through the plasma is considered
to lose the energy via two-body collisions with thermal quarks
or gluon emission.
Although it is detectable through jet-quenching effect at RHIC,
this energy loss rate cannot be calculated analytically
beyond the perturbative QCD.

Gauge/gravity duality~\cite{maldacena97,Witten98,GKP98}
shed some light on the investigation
of the strongly coupled gauge theories.
This duality states that the expectation value of
the energy-momentum tensor $\Exp{T_{\mu\nu}}$
of thermal ${\cal N}=4$ super-Yang-Mills (SYM) theory
at large 't Hooft coupling $g_{\text{YM}}^2 N$
and a large number of colors $N$ can be read off
from the perturbed metric of anti-de Sitter~(AdS) black hole.
For example, we can obtain transport coefficients such as
shear viscosity by calculating quasi-normal mode of AdS black hole.
Many calculations showed that the ratio of viscosity divided
by entropy density is equal to $1/4\pi$
for a large class of gauge theories described by gravity duals~
(See, for example, Ref.~\cite{KovSonStar04,mno06} for all references).
This universal value is in good agreement with the RHIC data,
which suggests that it is a universal feature
for all strongly coupled gauge theories, including QCD.
%This agreement leads us to speculate about universal properties
%of the strongly coupled plasmas and to investigate
%the energy dissipation via gauge/gravity duality.
This agreement leads us to seek other universal properties
of the strongly coupled plasmas.
Even though there is no universality,
we would obtain some insights
which are not readily obtained
from perturbative techniques and lattice simulations.

In gauge/gravity duality,
a heavy quark corresponds to an open string
which stretches from the boundary of AdS spacetime
to the horizon of the black hole.
Under the test string approximation,
a large number of studies have been made on estimating
a jet-quenching parameter or energy loss rate of the quark
moving through the plasma via gauge/gravity duality~%
\cite{HKKKY,gubser06,herzog06,LiuRajWie06,CaTe06,Buchel06,%
CacGui06,FGMP065}.
Beyond the test string approximation,
we need to take into account the back-reaction
by the radiation from the string onto the original AdS geometry,
as pointed out in~\cite{FGMP065}.
So, a good place to start is to investigate
the properties of radiation emitted by the string.
In a series of papers
\cite{GubserYarom0709,GPY07064,GPY07060,GubserPufu07,FGMP06},
the energy density and energy flux associated
with the quark with constant velocity were calculated
from the metric perturbation by the string trailing behind the quark.
In the case of supersonic motion, the Mach cone structure,
which is characteristic of Cherenkov radiation, was also displayed.

In this paper we shall investigate the properties of radiation
from an accelerated quark via gauge/gravity duality.
%In electromagnetic case, for exmaple, the radiation
%corresponds to Bremsstrahlung radiation by a scattered electron.
Since the quarks in the plasma would be scattered by one another,
the bremsstrahlung radiation by the scattered quark
would bring about another mechanism of energy dissipation
of the quark moving through the plasma.
Callan and G\"{u}ijosa~\cite{CallanGuijosa99} first investigated
time-dependent fields produced by an oscillating quark via gauge/gravity duality 
and showed that the one point function $\Exp{\mbox{Tr}F^2}$
in the boundary theory decays faster than $1/R^2$.
This does not immediately mean that there is no radiation 
emitted by the quark. As mentioned in~\cite{CallanGuijosa99,SinZahed05},
we can explain the reason for the curious behavior of the function
%does not decay as $1/R^2$ 
by analogy with electromagnetism as follows:
$F^2$ is expressed by the electric field $E$ and magnetic field $B$ as $F^2=E^2-B^2$. 
Since $E^2=B^2\sim 1/R^2$ for radiation field,
the leading radiation parts are cancelled in $F^2$. 
So, we show the existence of the radiation by dealing with energy-momentum 
tensor of the gauge theory generated by an accelerated quark. 

As a first step, we calculate energy density
or energy flux of radiation from an accelerated quark
at zero temperature by solving master equations
of linearized Einstein equations sourced by the string
given in~\cite{GubserPufu07}.
In section \ref{sec:review} we briefly review the master equation
and the general method to solve the equation
developed in~\cite{GubserPufu07}.
In section \ref{sec:oscillating}
we calculate the time averaged energy density $\Exp{T_{00}}$
of radiation in the case of an oscillating quark
with frequency $\Omega$.
%It is shown that the energy density is asymptotically istropic and 
%it falls off asymptotically as $\sqrt{g_{\text{YM}}^2 N}\,
%\Omega^4/|\bmx|^{2}$.   
It is shown that the energy density is asymptotically isotropic
and it falls off as $\sqrt{g_{\text{YM}}^2 N}~\Omega^4/R^{2}$.
In section \ref{sec:bremss} we simply estimate the Poynting vector
associated with the bremsstrahlung radiation
in a toy model of a scattered quark by an external field.
It is shown that the energy flux is anisotropic outgoing radiation.
Conclusion and discussions are devoted in the final section.
%%%%%%%%%%%%%%%%%%%%%%%%%%%%%%%%%%%%%%%%%%
\section{Master equation for a scalar mode}
\label{sec:review}
%%%%%%%%%%%%%%%%%%%%%%%%%%%%%%%%%%%%%%%%%% 
A quark in ${\cal N}=4$ SYM theory
at zero temperature corresponds to a fundamental string
through the Nambu-Goto action
\be
\label{nambugoto}
S=-\frac{1}{2\pi\alpha'}\int d\sigma^2\sqrt{-h},
\ee
where $h$ is the determinant of the world-sheet metric
$h_{\alpha\beta}=g_{\mu\nu}\p_\alpha X^\mu \p_\beta X^\nu$.
The background metric $g$ is given by the metric of $AdS_5$ spacetime
\begin{align}
\label{AdSmetric}
% & ds^2 = a(r)^2 \left( -dt^2 + dx_i^2 \right) + \frac{dr^2}{a(r)^2}~,
  & ds^2 = a(r)^2 \left( -dt^2 + dx_i^2 \right) + dr^2/a(r)^2~,
& & a(r) = r/L~, %\frac{r}{L}~,
& & (i=1,2,3)~,
\end{align}
where $L$ and $\alpha'$ are related
by 't Hooft coupling $g_{\mathrm{YM}}^2N$ as
\be 
%\frac{L^4}{\alpha'^2}=g_{YM}^2N~.
  L^4/\alpha'^2 = g_{\mathrm{YM}}^2N~.
\ee

We solve the linearized Einstein equations in the metric
sourced by the accelerated string.
Since we are interested in the energy density or energy flux
of the bremsstrahlung radiation,
we shall focus on the scalar perturbations
of the linearized Einstein equations.
Following~\cite{GubserPufu07},
the scalar modes of the string energy-momentum tensor
$\tau_{\mu\nu}$ and the metric perturbations $h_{\mu\nu}$
are expressed as
\begin{align}
  & \tau_{ab}(t,\bmx,r)
  = \int d\omega \int \frac{d^3k}{(2\pi)^3}~
    \tau^S_{bc}(\omega,\bmk,r)~ \bbS(\bmk,\bmx)~e^{-i\omega t}~,
\label{scalar-EM-def1} \\
  & \tau_{bi}(t,\bmx,r)
  = a(r)\int d\omega \int \frac{d^3k}{(2\pi)^3}~
  \left[~\tau^S_b(\omega,\bmk,r)~\bbS_i(\bmk,\bmx)~\right]~
  e^{-i\omega t}~,
\label{scalar-EM-def2} \\
  & \tau_{ij}(t,\bmx,r)
  = 2a^2(r)\int d\omega \int \frac{d^3k}{(2\pi)^3}~
  \left[~p^S(\omega,\bmk,r)~\delta_{ij}~\bbS(\bmk,\bmx)
    + \tau^S(\omega,\bmk,r)~\bbS_{ij}(\bmk,\bmx)~\right]~
  e^{-i\omega t}~,
\label{scalar-EM-def3}
\end{align}
and
\begin{align}
  & h_{ab}(t,\bmx,r)
  = \int d\omega \int \frac{d^3k}{(2\pi)^3}~f_{ab}^S(\omega,\bmk,r)~
    \bbS(\bmk,\bmx)~e^{-i\omega t}~,
\\
  & h_{bi}(t,\bmx,r)
  = a(r)\int d\omega \int \frac{d^3k}{(2\pi)^3}~
  \left[~f^S_b(\omega,\bmk,r)~\bbS_i(\bmk,\bmx)~\right]~
  e^{-i\omega t}~,
\\
  & h_{ij}(t,\bmx,r)
  = 2a^2(r)\int d\omega \int \frac{d^3k}{(2\pi)^3}~
  \left[~H^S_L(\omega,\bmk,r)~\delta_{ij}~\bbS(\bmk,\bmx)
  + H^S_T(\omega,\bmk,r)~\bbS_{ij}(\bmk,\bmx)~\right]~e^{-i\omega t}~,
\end{align}
%
%\be
%\label{scalar-EM-def1}
%\tau_{ab}(t,\bmx,r)=\int d\omega \int \frac{d^3k}{(2\pi)^3}~\tau^S_{bc}(\omega,\bmk,r)~
% \bbS(\bmk,\bmx)~e^{-i\omega t}~,
%\ee
%\be 
%\label{scalar-EM-def2}
%\tau_{bi}(t,\bmx,r)=a(r)\int d\omega \int \frac{d^3k}{(2\pi)^3}~
%\left[~\tau^S_b(\omega,\bmk,r)~\bbS_i(\bmk,\bmx)~\right]~ e^{-i\omega t}~,
%\ee 
%\be 
%\label{scalar-EM-def3}
%\tau_{ij}(t,\bmx,r)&=&2a^2(r)\int d\omega \int \frac{d^3k}{(2\pi)^3}~
%\left[~p^S(\omega,\bmk,r)~\delta_{ij}~\bbS(\bmk,\bmx)+\tau^S(\omega,\bmk,r)~\bbS_{ij}(\bmk,\bmx)~\right]~
%e^{-i\omega t}~, 
%\ee
%\be 
%h_{ab}(t,\bmx,r)=\int d\omega \int \frac{d^3k}{(2\pi)^3}~
%f_{ab}^S(\omega,\bmk,r)~\bbS(\bmk,\bmx)~e^{-i\omega t}~,
%\ee
%\be 
%h_{bi}(t,\bmx,r)=a(r)\int d\omega \int \frac{d^3k}{(2\pi)^3}~
%\left[~f^S_b(\omega,\bmk,r)~\bbS_i(\bmk,\bmx)~\right]~e^{-i\omega t}~,
%\ee 
%\be 
%h_{ij}(t,\bmx,r)&=&2a^2(r)\int d\omega \int \frac{d^3k}{(2\pi)^3}~
%\left[~H^S_L(\omega,\bmk,r)~\delta_{ij}~\bbS(\bmk,\bmx)+
%H^S_T(\omega,\bmk,r)~\bbS_{ij}(\bmk,\bmx)~\right]~e^{-i\omega t}~,
%\ee
where $a, b = t, r$ and
functions $\bbS$, $\bbS_i$, and $\bbS_{ij}$ are defined as
\begin{align}
  & \bbS(\bmk,\bmx) = e^{i\bmk\cdot\bmx}~,
& & \bbS_i(\bmk,\bmx)=-\frac{1}{k}\p_i\bbS(\bmk,\bmx)~,
\nonumber \\
  & \bbS_{ij}(\bmk,\bmx) = \frac{1}{k^2}~\p_i\p_j~\bbS(\bmk,\bmx)
+\frac{1}{3}~\delta_{ij}~\bbS(\bmk,\bmx)~.
\end{align}
Here $k$ is defined by $k := |\bmk |$.

The diffeomorphism-invariant quantity of $f^S_{ab}$ is constructed 
as~\cite{GubserPufu07}
\begin{align} 
  & \hat{f}^S_{ab} = f^S_{ab} + 2 D_{(a}X_{b)}~,
& & X_a
  = \frac{a(r)}{k}\,
  \left( f^S_a + \frac{a(r)}{k}\, \p_a H^S_T \right)~.
\end{align} 
Defining the master field%
\footnote{We set $\kappa_5^2=1$.} as
\be 
\Phi(\omega,\bmk,r)
= - \frac{k^2}{2a^3L}\, \left( \frac{\hat{f}^S_{tt}}{a}
- a^3 \hat{f}^S_{rr} + \frac{a^3\hat{f}^S_{tr}}{ir\omega} \right)~,
\ee
the master equation is given by 
\be 
\label{master-eq}
\mathcal{L}\, \Phi(\omega,\bmk,r)
:= \left(\p_r^2 + \frac{5}{r}\p_r - \frac{L^4}{r^4}(k^2-\omega^2)
+ \frac{4}{r^2}\right) \Phi(\omega,\bmk,r)
= - j(\omega,\bmk,r)~,
\ee
with a source term of the string
\begin{align}
\label{st-source}
   j(\omega,\bmk,r)
  &= \frac{1}{3 L^3 \omega r^3}\, \Big[~
  k^2 L^2 \omega\, r^3 \tau_{rr}(\omega,\bmk,r)
  + 2 \omega (3 L^4 \omega^2\, r - 4 k^2 L^4\, r + 48 r^3)
    \tau^S(\omega,\bmk,r)
\nonumber \\
  &\hspace*{1.5cm}
  - i k L \big\{ - 3 i \omega\, r^3 \tau^S_r(\omega,\bmk,r)
  + ( k^2 L^4 - 9 r^2 ) \tau^S_t(\omega,\bmk,r) \big\}~\Big]
\nonumber \\
  &- \frac{i}{3 L^3 \omega\, r}\,
  \Big[~2 k^2 L^2 \tau^S_{tr}(\omega,\bmk,r)
  + k L\, r \left\{ k L \tau'^S_{tr}(\omega,\bmk,r)
  - 3 \tau'^S_t(\omega,\bmk,r) \right\}
\nonumber \\
  &\hspace*{1.5cm}
  + 6 i r^2 \omega \left\{ 9 \tau'^S(\omega,\bmk,r)
  + r\, \tau''^S(\omega,\bmk,r) \right\}~\Big]~,
\end{align}
where a prime means the derivative with respect to $r$.

Eq.~(\ref{master-eq}) can be solved
by a Green function $G_R(\omega,\bmk,r;r_0)$ which obeys
\be 
%\left(\p_r^2+\frac{5}{r}\p_r-\frac{L^4}{r^4}(k^2-\omega^2)
%+\frac{4}{r^2}\right) 
{\cal L}\,G_R(\omega,\bmk,r;r_0)
= \delta(r-r_0)~.
\ee
This is constructed by two solutions of the homogeneous equation
$\mathcal{L} \Phi_H = 0 = \mathcal{L} \Phi_B$,
where $\Phi_H$ satisfies appropriate boundary conditions
at the degenerate horizon $r=0$,
while $\Phi_B$ has appropriate falloff
at the boundary of $\text{AdS}_5$.
Then the Green function $G_R$ satisfying the boundary conditions
at both the horizon and infinity 
%satisfied by $\Phi_H$ and $\Phi_B$, respectively, 
is represented by $\Phi_B$ and $\Phi_H$ as
\be 
  G_R(\omega,\bmk,r;r_0)
= \frac{1}{W[\Phi_H,\Phi_B](r_0)}~\Big[~
    \Phi_H(r_0) \Phi_B(r) \theta(r-r_0)
  + \Phi_H(r) \Phi_B(r_0) \theta(r_0-r)~\Big]~,
\ee
where $W[\Phi_H,\Phi_B](r_0)$ is the Wronskian defined by
\be
W[\Phi_H,\Phi_B](r)=\Phi_H(r)\p_r\Phi_B(r)
-\Phi_B(r)\p_r\Phi_H(r)~.
\ee
So, the solution of Eq.~(\ref{master-eq}) is given by
\be
\label{r-solution}
\Phi(\omega,\bmk,r)=A(r)\Phi_H(r)+B(r)\Phi_B(r)~
\ee
with
\begin{align}
\label{coefficient}
  & A(r)=-\int^\infty_r dr_0~\frac{\Phi_B(r_0)}{W(r_0)}~j(r_0)~,
& & B(r)=-\int^r_0 dr_0~\frac{\Phi_H(r_0)}{W(r_0)}~j(r_0)~.
\end{align}

As discussed in~\cite{FGMP06},
the expectation value $\Exp{T_{\mu\nu}(t, \bmx)}$ of the energy-momentum tensor of 
${\cal N}=4$ SYM theory can be read off from the asymptotic behavior of $\Phi(\omega,\bmk,r)$
near $\text{AdS}_5$ boundary,
\be 
\label{asymptotics}
\Phi(\omega, \bmk, r)
= \frac{P(\omega, \bmk)}{r}+\frac{Q(\omega, \bmk)}{r^2}+O(r^{-3})~.
\label{eq:falloff}
\ee
The term $P/r$ in Eq.(\ref{eq:falloff}) produces singularity
of $\Exp{T_{\mu\nu}(t, \bmx)}$ supported at the location of the quark,
% to the energy-momentum tensor,
while the $Q/r^2$ term gives the energy density of the radiation as
\begin{align}
  & \Exp{T_{00}(\omega, \bmk)} = Q(\omega, \bmk)~.
\end{align}
%
%Let us suppose that the brane is located at $r=R~(>>1)$. Since the endpoint of 
%the string lies on the brane, the master field $\Phi$ at $r=R$ 
%is simply given by 
%\be 
%\Phi(\omega,\bmk,R)=B(R)\Phi_B(R).  
%\ee
%As discussed in~\cite{FGMP06}, we can read $\Exp{T_{\mu\nu}}$ from the asymptotic fall 
%off of $B(R)$.  
%%%%%%%%%%%%%%%%%%%%%%%%%%%%%%%%
\section{radiation from an Oscillating String}
\label{sec:oscillating}
%%%%%%%%%%%%%%%%%%%%%%%%%%%%%%%%
In this section, we calculate the energy density
$\Exp{T_{00}}$ of ${\cal N}=4$ SYM radiation
from an oscillating quark.
This can be obtained by solving Eq.~(\ref{master-eq})
for an oscillating open string. 

Now we consider that the string oscillates only
in the $x_1$ direction~($X_2=X_3=0$). 
Under the static gauge $\tau=t$, \, $\sigma=z=L^2/r$,
the equation of motion is obtained from Eq.~(\ref{nambugoto}) as
\be
z^2\, \frac{\p}{\p z} \left(
\frac{X'_1}{ z^2 \sqrt{1 + {X'_1}^2 - \dot{X}_1^2} } \right)
- \frac{\p}{\p t} \left(
\frac{z^2\, \dot{X}_1}{ \sqrt{1 + {X'_1}^2 - \dot{X}_1^2} } \right)
= 0~.
\ee
Here a dot and a prime mean the derivatives
with respect to $t$ and $z$, respectively.
Assuming that the oscillations are very small~($|X_1| \ll 1$),
the linearized equation is simply given by
\be
\label{linear-motion}
X_1''-\frac{2}{z}X'_1-\ddot{X_1}=0~.
\ee
Under the ansatz $X_1(t,z)=\sin(\Omega t) \xi(z)$,
it is easily checked that
\begin{align}
\label{solution-os-string}
  & \xi(z)
  = \epsilon \left\{ \cos(\Omega z) + \Omega z\sin(\Omega z) \right\}~,
& & z = L^2/r~,
\end{align}
is a solution of Eq.~(\ref{linear-motion}),
where $\epsilon$ is a small parameter and
$\Omega$ is a positive constant,
representing the angular frequency of the oscillating string.
Calculating the energy-momentum tensor $\tau^{\mu\nu}$,
\be
\tau^{\mu\nu} = \frac{2}{\sqrt{-g}}~\frac{\delta S}{\delta g_{\mu\nu}}
= - \frac{1}{2\pi\alpha'}~\frac{\sqrt{-h}}{\sqrt{-g}}~
h^{\alpha\beta} (\p_\alpha X^\mu) (\p_\beta X^\nu)\, \delta(x^1-X^1)\,
\delta(x^2)\, \delta(x^3)~,
\ee
one gets the non-zero components up to the second order
\begin{align}
   \tau^{tt}
  &= \frac{1}{4\pi\alpha' a^5(r)}~\int \frac{d^3k}{(2\pi)^3}~
  e^{i \bmk\cdot \bmx}~\Big( 2 - 2\, i\, k_1\, \sin(\Omega t)~\xi(r)
  + \Omega^2\, \cos^2(\Omega t)~\xi^2(r)
\nonumber \\
  &\hspace{4.5cm}
  + a^4(r)\, \sin^2(\Omega t)~\xi'^2(r)
  - k_1^2\, \sin^2(\Omega t)~\xi^2(r) \Big)~,
\nonumber \\
   \tau^{rr}
  &= \frac{-1}{4\pi\alpha' a(r)}~\int \frac{d^3k}{(2\pi)^3}~
  e^{i \bmk\cdot \bmx}~\Big( 2 + 2\, i\, k_1\, \sin(\Omega t)~\xi(r)
  - \Omega^2\, \cos^2(\Omega t)~\xi^2(r)
\nonumber \\
  &\hspace{4.5cm}
  - a^4(r)\, \sin^2(\Omega t)~\xi'^2(r)
  - k_1^2\, \sin^2(\Omega t)~\xi^2(r) \Big)~,
\nonumber \\
   \tau^{rt}
  &= - \frac{\Omega}{2\pi\alpha'a(r)}~
  \cos(\Omega t)\, \sin(\Omega t)~\xi(r)\, \xi'(r)
  \int \frac{d^3k}{(2\pi)^3}~e^{i \bmk\cdot \bmx}~,
\nonumber \\
   \tau^{t1}
  &= \frac{\Omega}{2\pi\alpha'a^5(r)}~\cos(\Omega t)~\xi(r)\,
  \Big( 1 - i\, k_1\, \sin(\Omega t)~\xi(r) \Big)\,
  \int \frac{d^3k}{(2\pi)^3}~e^{i \bmk\cdot \bmx}~,
\nonumber \\
   \tau^{r1}
  &= - \frac{1}{2\pi\alpha'a(r)}~\sin(\Omega t)~\xi'(r)
  \Big( 1 - i\, k_1\, \sin(\Omega t)~\xi(r) \Big)\,
  \int \frac{d^3k}{(2\pi)^3}~e^{i \bmk\cdot \bmx}~,
\nonumber \\
   \tau^{11}
  &= \frac{1}{2\pi\alpha' a^5(r)}~
  \Big( \Omega^2\, \cos^2(\Omega t)~\xi^2(r)
  - a^4(r)\, \sin^2(\Omega t)~\xi'(r)^2 \Big)\,
  \int \frac{d^3k}{(2\pi)^3}~e^{i \bmk\cdot \bmx}~,
%\tau&=&e^{-i\bmk\cdot \bmx}
\end{align}
%
%\be 
%\tau^{tt}&=&\frac{1}{4\pi\alpha' a^5(r)}
%\int \frac{d^3k}{(2\pi)^3}\,e^{i\bmk\cdot \bmx}
%%\{2-2i\sin(\Omega t)k_1\xi(r)
%+\Omega^2\cos^2(\Omega t)\xi^2(r) \nonumber \\
%&&\hspace{3cm}+a^4(r)\sin^2(\Omega t)\xi'^2(r)-k_1^2\sin^2(\Omega t)\xi^2(r)\},
% \nonumber \\
%\tau^{rr}&=&\frac{-1}{4\pi\alpha' a(r)}
%\int \frac{d^3k}{(2\pi)^3}\,e^{i\bmk\cdot \bmx}
%\{2+2i\sin(\Omega t)k_1\xi(r)-\Omega^2\cos^2(\Omega t)\xi^2(r) \nonumber \\
%&&\hspace{3cm}-a^4(r)\sin^2(\Omega t)\xi'^2(r)-k_1^2\sin^2(\Omega t)\xi^2(r)\},
% \nonumber \\
%\tau^{rt}&=&-\frac{\Omega}{2\pi\alpha'a(r)}
%\cos(\Omega t)\sin(\Omega t)\xi(r)\xi'(r)
%\int \frac{d^3k}{(2\pi)^3}\, e^{i\bmk\cdot \bmx}, \nonumber \\
%\tau^{t1}&=&\frac{\Omega}{2\pi\alpha'a^5(r)}\cos(\Omega t)\xi(r)
%\{1-i\sin(\Omega t)k_1\xi(r)\}
%\,\int \frac{d^3k}{(2\pi)^3}\,e^{i\bmk\cdot \bmx},
% \nonumber \\
%\tau^{r1}&=&-\frac{1}{2\pi\alpha'a(r)}\sin(\Omega t)\xi'(r)
%\{1-i\sin(\Omega t)k_1\xi(r)\}
%\,\int \frac{d^3k}{(2\pi)^3}\, e^{i\bmk\cdot \bmx},
% \nonumber \\
%\tau^{11}&=&\frac{1}{2\pi\alpha' a^5(r)}
%\{\Omega^2\cos^2(\Omega t)\xi^2(r)-a^4(r)\sin^2(\Omega t)\xi'(r)^2\}
%\,\int \frac{d^3k}{(2\pi)^3}\,e^{i\bmk\cdot \bmx}, \nonumber \\
%%\tau&=&e^{-i\bmk\cdot \bmx}
%\ee
where a dash denotes the derivative with respect to $r$.
Substituting these into Eqs.~(\ref{scalar-EM-def1})%
-(\ref{scalar-EM-def3}), the components $\tau^S$, $P^S$,
$\tau^S_t$, $\tau^S_r$, $\tau^S_{rr}$, and $\tau^S_{tr}$ are
\begin{align}
\label{scalar-coe}
   \tau_t^S
  &= - \frac{i\, \Omega\, k_1\, \xi}{4 \pi \alpha'~k\, a^2(r)}\,
  \left[~\delta(\omega+\Omega)+\delta(\omega-\Omega)
  - k_1\, \xi~\frac{\delta(\omega+2\Omega)-\delta(\omega-2\Omega)}{2}
  ~\right]~,
\nonumber \\
   \tau_r^S
  &= - \frac{k_1\, \xi'}{4 \pi \alpha'~k\, a^2(r)}\,
  \left[~\delta(\omega+\Omega) - \delta(\omega-\Omega)
    + k_1\, \xi\, \delta(\omega)
  - k_1\, \xi~\frac{\delta(\omega+2\Omega)+\delta(\omega-2\Omega)}{2}\,
  ~\right]~,
\nonumber \\
   P^S
  &= \frac{1}{24 \pi \alpha'~a(r)}\,
  \Biggl[~\left( \frac{\Omega^2}{a^2(r)}\, \xi^2 -a^2(r)\, \xi'^2
          \right) \delta(\omega)
%\nonumber \\
  + \left( \frac{\Omega^2}{a^2(r)}\, \xi^2 + a^2(r)\, \xi'^2 \right)
  \frac{ \delta(\omega+2\Omega) +\delta(\omega-2\Omega) }{2}~\Biggr]~,
\nonumber \\
   \tau^S
  &= - \frac{3}{2}~\left( \frac{3k_1^2}{k^2}-1 \right)~P^S~,
\nonumber \\
%   \tau^S
%  &= -\frac{1}{16\pi \alpha'a(r)}\, \left( \frac{3k_1^2}{k^2}-1 \right)
%  \Biggl[~\left( \frac{\Omega^2}{a^2(r)}\, \xi^2 - a^2(r) \xi'^2 \right)
%  \delta(\omega)
%%\nonumber \\
%  + \frac{1}{2}\, ( \delta(\omega+2\Omega) + \delta(\omega-2\Omega) )
%  \left( \frac{\Omega^2}{a^2(r)}\, \xi^2 + a^2(r) \xi'^2 \right)~
%  \Biggr]~,
%\nonumber \\
   \tau^S_{rr}
  &= \frac{1}{8 \pi \alpha'~a^5(r)}\, \Bigg[~
   \big( a^4(r) \xi'^2 + k_1^2 \xi^2 + \Omega^2 \xi^2 - 4 \big)~
   \delta(\omega)
%\nonumber \\
  + k_1\xi\, \frac{\delta(\omega+\Omega) - \delta(\omega-\Omega)}{2}
\nonumber \\
  &\hspace*{2.0cm}
  + \big( \Omega^2 \xi^2 - a^4(r) \xi'^2 - k_1^2 \xi^2 \big)~
  \frac{\delta(\omega+2\Omega) + \delta(\omega-2\Omega)}{2}~\Bigg]~,
\nonumber \\
   \tau^S_{tr}
  &= - \frac{i\, \Omega}{8 \pi \alpha'~a(r)}~\xi\, \xi'~
  \big( \delta(\omega+2\Omega) - \delta(\omega-2\Omega) \big)~.
\end{align}
We are interested in the time integral of the energy density,
\begin{align}
  & \int^{\infty}_{-\infty} dt~\Exp{T_{00}(t, \bmx)}
  = \int^{\infty}_{-\infty} dt \int^{\infty}_{-\infty} d\omega~
      \Exp{T_{00}(\omega, \bmx)}~e^{-i \omega t}
  = 2\, \pi~\Exp{T_{00}(\omega=0, \bmx)}~.
\label{eq:total_E}
\end{align}
Owing to the linearity of Eq.(\ref{master-eq}),
only the terms with $\delta(\omega)$ in Eq.(\ref{scalar-coe})
contribute to the time integral of the energy density. 
%we drop the terms except the terms with
%$\delta(\omega)$in Eq.(\ref{scalar-coe}).
So, we drop the terms with $\delta(\omega \pm \Omega)$
or $\delta(\omega \pm 2 \Omega)$ in Eq.(\ref{scalar-coe}).
Substituting Eq.~(\ref{solution-os-string})
and Eq.~(\ref{scalar-coe}) into the source term~(\ref{st-source}),
one obtains
\begin{align}
j(\omega, \bmk, r) &= j_0(\bmk, r)~\delta(\omega) + \cdots~,
\\
j_0(\bmk, r)&=\epsilon^2\left(\frac{C_1}{r^3}+\frac{C_2}{r^5}
+\frac{C_3}{r^7}   \right)\cos(2\Omega z)
\nonumber \\
&+\epsilon^2\left(\frac{S_1}{r^4}+\frac{S_2}{r^6}   \right)\sin(2\Omega z)
\nonumber \\
&+\epsilon^2\left(\frac{N_1}{r^3}+\frac{N_2}{r^5}+\frac{N_3}{r^7} \right)~,
\end{align}
where all the coefficients are given by
\be 
  C_1&=& \frac{\Omega^2}{16 \pi \alpha'}\,
  \left( 1 - \frac{3k_1^2}{k^2} \right)~,
\qquad 
  C_2= \frac{L^4}{48 \pi \alpha'}\, \left(
  18 \Omega^4\, \frac{k_1^2}{k^2} + 6 \Omega^2 (k_1^2-\Omega^2)
  + k^2 (k_1^2-3\Omega^2) \right)~,
\nonumber \\
  C_3&=& \frac{L^8\, \Omega^2}{48 \pi \alpha'}\, \left[~
  - k^2 k_1^2 + 8 \Omega^2 ( k^2 + 3 \Omega^2)
      \left( 1 - \frac{3 k_1^2}{k^2} \right)~
  \right]~,
\nonumber \\
  S_1&=& \frac{L^2\, \Omega^3}{8 \pi \alpha'}\,
  \left( 1 - \frac{3k_1^2}{k^2} \right)~,
\qquad
  S_2= \frac{L^6\, \Omega}{24 \pi \alpha'}\, \left(
  k^2 k_1^2 - 3\Omega^2(k^2-3 k_1^2) + 6\frac{\Omega^4}{k^2}(k^2-3k_1^2) \right)~,
\nonumber \\
  N_1&=& \frac{\Omega^2}{16 \pi \alpha'}\,
  \left( 1 - \frac{3k_1^2}{k^2} \right)~,
\qquad
  N_2= \frac{L^4 k^2}{48 \pi \alpha'}\, \left(
  6\Omega^2\, \frac{k_1^2}{k^2} + k_1^2 - 3 \Omega^2 \right)~,
\nonumber \\
  N_3&=& \frac{L^8\, \Omega^2}{48 \pi \alpha'}~
  k^2\, (k_1^2+2\Omega^2)~.
\ee
The two-independent homogeneous solutions $\Phi_H$ and $\Phi_B$
for $\omega=0$ are simply represented
by the modified Bessel function as
\begin{align}
  & \Phi_B(r)=\frac{I_0(kL^2/r)}{r^2}~,
%& & \Phi_H(r)=\frac{K_0(kL^2/r)}{r^2}~.
& & \Phi_H(r)=\frac{K_0(kL^2/r)}{r^2}~,
\end{align}
and their Wronskian is simply given by
%The Wronskian is simply given by
\be 
W = - 1/r^5~.
\ee
%The asymptotic solution~(\ref{r-solution}) behaves as  
%\be 
%\Phi(r)=\frac{\alpha}{r}+\frac{\beta}{r^2}+O(r^{-3}). 
%\ee 
%According to gauge/gravity duality, the expectation value of the energy 
%density of the energy-momentum tensor of the gauge field $\Exp{T_{00}}$ can be read 
%off from the coefficient $\beta$.

We can obtain the coefficient $Q$ in Eq.(\ref{eq:falloff})
through a straightforward but tedious calculation~
(the detailed calculation is given in the Appendix A). 
The coefficient $Q$ is given by 
\begin{align}
   Q(\omega, \bmk)
  &= Q_0(\bmk)~\delta(\omega) + \cdots~,
\\
   Q_0(\bmk)
  &= \frac{\pi\, \epsilon^2}{2 L^2\, \Omega}\,
     \frac{C_2}{ (4 + k^2/\Omega^2)^{1/2} }
  + \frac{3 \pi\, \epsilon^2}{2 L^6\, \Omega^5}\,
    \frac{C_3\, k^2}{ ( 4 + k^2/\Omega^2 )^{5/2} }
  - \frac{\pi\, \epsilon^2}{L^6\, \Omega^3}\,
    \frac{C_3}{ (4 + k^2/\Omega^2 )^{3/2} }
\nonumber \\
  &+ \frac{\pi\, \epsilon^2}{2}\, S_1
  \ln\left( \frac{2 \Omega}{k} + \sqrt{\frac{4\Omega^2}{k^2}+1} \right)
  + \frac{\pi\, \epsilon^2}{L^4}\,
    \frac{S_2\Omega}{ (k^2 + 4\Omega^2 )^{3/2} }
\nonumber \\
  &+ \frac{\pi\, \epsilon^2}{2}\, \left(
     k L^2 N_1 + \frac{N_2}{k L^2} + \frac{N_3}{(k L^2)^3} \right)~.
\end{align}
The expectation value $\Exp{T_{00}(\omega, \bmx)}$
in position space $\bmx$ is obtained
by the Fourier transformation,
\be 
\label{fourier}
\Exp{T_{00}(\omega, \bmx)} =
\int^\infty_{-\infty} \frac{d^3k}{(2 \pi)^3}~Q(\omega, \bmk)~
e^{i \bmk \cdot \bmx}~.
\ee
So, in the far asymptotic region $|\bmx| \gg 1$,
the integration is dominant for $k\to 0$.
In this limit, 
\be 
\label{coefficientB-asy}
\lim_{k\to 0}Q_0&\simeq&
\frac{\pi\, \epsilon^2}{4 L^2\, \Omega}
\left( C_2 + \frac{3 C_3\, k^2}{16 L^4\, \Omega^4}
- \frac{C_3}{2 L^4\, \Omega^2} \right)
\nonumber \\
&-&\frac{\pi\, \epsilon^2}{2}\, S_1\, \ln k
+\frac{\pi\, \epsilon^2}{8L^4\, \Omega^2}\, S_2
\nonumber \\
&+&\frac{\pi\, \epsilon^2}{2}\,
\left( k L^2 N_1 + \frac{N_2}{k L^2} + \frac{N_3}{(k L^2)^3} \right)~.
\ee
The integral can be easily performed
by differentiating appropriately modified versions
of the standard formulas
\begin{align}
  & \int \frac{d^3 k}{(2\pi)^3}~\frac{e^{i\bmk\cdot \bmx}}{|\bmk|}
  = \frac{1}{2\pi^2\, R^2}~,
& & \int \frac{d^3 k}{(2\pi)^3}~\frac{e^{i\bmk\cdot \bmx}}{|\bmk|^2}
  = \frac{1}{4\pi\, R}~,
\nonumber \\
  & \int \frac{d^3 k}{(2\pi)^3}~\frac{e^{i\bmk\cdot \bmx}}{|\bmk|^3}
  = -\frac{\ln R}{2\pi^2}~,
& & \int \frac{d^3 k}{(2\pi)^3}~\frac{e^{i\bmk\cdot \bmx}}{|\bmk|^4}
  = -\frac{\pi\, R}{8\pi}~, 
\end{align}
where $R := |\bmx |$.
So, Fourier transform of every term except last term
in Eq.~(\ref{coefficientB-asy}) decays faster than $1/R^{2}$~
\footnote{$\ln k$ term in Eq.~(\ref{coefficientB-asy}) also decays
faster than $1/R^{2}$.}.
The Fourier transform of the last term becomes
\be 
\label{density}
\Exp{T_{00}(\omega, \bmx)}
\sim \frac{\epsilon^2L^2\Omega^4}{96\pi^2\alpha'\, R^2}~\delta(\omega)
\sim \frac{\epsilon^2\Omega^4\sqrt{g_{\text{YM}}^2N}}{96\pi^2\, R^2}~
\delta(\omega). 
\ee
Substituting this into Eq.(\ref{eq:total_E}), we obtain
\begin{align}
  & \int^{\infty}_{-\infty} dt~\Exp{T_{00}(t, \bmx)}
%  = 2\, \pi~\Exp{T_{00}(\omega=0, \bmx)}
  \sim \frac{\epsilon^2\Omega^4\sqrt{g_{\text{YM}}^2N}}{96\pi^2\, R^2}~
  \big( 2 \pi\, \delta(0) \big)
  = \frac{\epsilon^2\Omega^4\sqrt{g_{\text{YM}}^2N}}{96\pi^2\, R^2}~
  \int^{\infty}_{-\infty} dt~.
\label{eq:average_E}
\end{align}
This relation implies that the coefficient of $\delta(\omega)$
in Eq.(\ref{density}) corresponds to the time averaged energy density
of the radiation.
The energy density $\Exp{T_{00}}$ can be expressed by the acceleration,
$\Dot{v} \sim \epsilon\Omega^2$,
as $\sqrt{g_{\text{YM}}^2N}~\Dot{v}^2/R^2$.

Let us compare Eq.~(\ref{density})
with the times averaged energy density of electromagnetic radiation
sourced by an oscillating electron along $x_1$ axis
with the same amplitude $\epsilon$ and frequency $\Omega$.
In terms of charge $e$ of the electron, 
%permittivity $\epsilon_0$, and the speed of light $c$,
% the zero mode of the energy density $\rho$ is asymptotically
%written by
the power radiated per unit solid angle becomes
\be 
%\rho(\omega, \bmx)=\frac{e^2\epsilon^2\Omega^4}{32\pi^2\epsilon_0^2c^4|\bmx|^2}
%\sin^2\Theta, 
%\frac{dP}{d\Omega_{\mathrm{angle}}} = 
\frac{e^2\epsilon^2\Omega^4}{32\pi^2\, R^2}~\sin^2\Theta
\propto \frac{e^2 \Dot{v}^2}{R^2}~\sin^2\Theta~,
\label{eq:EM_dipole}
\ee
where $\Theta$ is the angle between $x_1$ axis and $\bmx$.
Then the time averaged energy density
is obtained by Eq.(\ref{eq:EM_dipole}) divided by the speed of light.

So, both of strongly coupled gauge field
and electromagnetic radiation densities agree
on the asymptotic falloff and the acceleration dependence,
while the angular distribution and the coupling dependence
in the energy density are quite different.
As seen in the case of a static quark~\cite{DanKesKruc},
the energy density in Eq.~(\ref{density}) is proportional
to the square of 't Hooft coupling, $\sqrt{g_{\text{YM}}^2N}$,
which is characteristic of strong coupling.
We will argue on the angular distribution in Sec \ref{sec:conclusion}.

To investigate the energy loss rate of the accelerated quark,
we need to obtain Poynting vector $\Exp{T^{0i}}$.
Unfortunately, it is technically difficult to explicitly calculate
$\Exp{T^{0i}}$ because of the complicated source~(\ref{st-source})
for the oscillating string solution~(\ref{solution-os-string}).
So, in the next section, we simply estimate the Poynting vector
$\Exp{T^{0i}}$ by using a toy model where
a straight string which moves at constant velocity
and then suddenly stops.
%%%%%%%%%%%%%%%%%%%%%%%%%%%%%%%
\section{Bremsstrahlung radiation from a scattered quark}
\label{sec:bremss}
%%%%%%%%%%%%%%%%%%%%%%%%%%%%%%%
In this section, we shall consider bremsstrahlung radiation
from a string which moves at constant velocity $v$ along $x_1$-axis,
suddenly stops at the origin and then remains at rest.
The Fourier transform $j(t,\bmk,r)$
of the source $j(\omega,\bmk, r)$ in Eq.~(\ref{master-eq})
can be constructed by the source term $J(t,\bmk, r;v)$
of the string moving at constant velocity $v$ for $-\infty<t<\infty$ as
\be 
%\hat{j}(t,\bmk,r)=\hat{J}(t,\bmk,r;v)\theta(-t)
%+\hat{J}(t,\bmk,r;0)\theta(t), 
j(t,\bmk,r) = J(t,\bmk,r;v)\theta(-t) + J(t,\bmk,r;0)\theta(t)~,
\ee
where the time when the string stops is set to $0$.
%where $\theta(t)$ is a step function and given by 
%\be 
%\theta(t)=\lim_{\epsilon\to +0}
%\int^\infty_{-\infty} \frac{du}{2\pi i}~\frac{e^{itu}}{u-i\epsilon}~.   
%\ee

So, using Fourier transform $j(\omega,\bmk,r)$ of $j(t,\bmk,r)$,
we rewrite the master equation in Eq.~(\ref{master-eq}) as
\begin{align}
& {\cal L}\Phi 
%&=& \left(\p_r^2+\frac{5}{r}\p_r+\frac{L^2}{r^4}(\omega^2-k^2)+\frac{4}{r^2}\right)\Phi \nonumber \\
= - j(\omega,\bmk,r) %\nonumber \\
= - i \left[~\frac{J(\bmk,r;v)}{\omega_--k_1v}
  - \frac{J(\bmk,r;0)}{\omega_+}~\right]~,
& & \omega_{\pm}=\omega\pm i\epsilon~,
\end{align}
where the source term of the string moving at constant velocity $v$
is given by~\cite{GubserPufu07}
\begin{align}
   J(\bmk,r;v)
  &= \frac{l(v)}{6 k^2}\, \left\{ k^2 (2+v^2) - 3 v^2 k_1^2 \right\}
  \left( - \frac{3}{r^3} + \frac{L^4}{r^5}\, (2k^2-3v^2k_1^2) \right)~,
\nonumber \\
   l(v)
  &= \frac{1}{2\pi\alpha'}\, \frac{1}{\sqrt{1-v^2}}~.
\end{align}
In order to construct two independent solutions $\Phi_H$, $\Phi_B$
which satisfy the homogeneous equation $\mathcal{L} \Phi = 0$
as well as the boundary conditions at the horizon and infinity,
we introduce \lq\lq radial wave number\rq\rq\,  in the bulk,
$q := \big( \omega^2 - k^2 \big)^{1/2}$.
For the multivalued function $q$ in the complex $\omega$-plane,
we take the branch cut such that $q$ is holomorphic in the upper plane,
that is,
\be 
%q=\left\{
%    \begin{array}{@{\,}ll} 
%         \sqrt{\omega^2-k^2}   
% & \mbox{for} \quad \omega>|\bmk| \\
%         i\sqrt{k^2-\omega^2} 
% &  \mbox{for} \quad |\bmk|>\omega>-|\bmk| \\         
%         -\sqrt{\omega^2-k^2} 
% &  \mbox{for} \quad -|\bmk|>\omega \\
%    \end{array} \right.~.
q = \begin{cases}
         \sqrt{\omega^2-k^2} & \text{~~for~~} \omega>k
      \\
         i\, \sqrt{k^2-\omega^2} & \text{~~for~~} k>\omega>-k
      \\         
         -\sqrt{\omega^2-k^2} & \text{~~for~~} -k>\omega
    \end{cases}~.
\ee
Then $\Phi_H$ and $\Phi_B$ are given by the Bessel functions as
\begin{align}
  & \Phi_B(r)=\frac{1}{r^2}J_0(qL^2/r)~,
& & \Phi_H(r)=\frac{1}{r^2}H^{(1)}_0(qL^2/r)~,
& & W=\frac{2i}{\pi r^5}~.  
\end{align}
Following to the procedure in the previous section,
we obtain the coefficient $Q$ in Eq.(\ref{eq:falloff}) as
\be 
\label{suddenly-coe}
   Q
  = -\frac{L^2}{8\alpha'k^2} \left[~
  \frac{ \left\{ 3v^2k_1^2-k^2(2+v^2) \right\}
  (k^2+3v^2k_1^2-3\omega^2) }{ 3 q (\omega_- - k_1 v) \sqrt{1-v^2} }
  + \frac{ 2 k^2 ( k^2 - 3 \omega^2 ) }{3 \omega_ + q}~\right]~.
%+O(r\ln r).  
\ee 
%The last divergent term cancels with the divergent term in the coefficient $A$.  
Supposing that $|v| \ll 1$, we shall expand $Q$ in terms of $v$ as 
\begin{align} 
\label{qs}
  Q &= (\text{terms up to $v$})
\nonumber \\
   &+ v^2~\frac{L^2}{\alpha'}~
   \frac{ 2 k^2 k_1^2 (k^2-3\omega^2)
     + \big\{ 2 k^4 + 9k_1^2 \omega^2 + 3 k^2 (k_1^2-2\omega^2) \big\}
     \omega_-^2}{24\, k^2\, \omega_-^3\, q} + O(v^3)~.  
\end{align}
The expansion would be valid for $|\omega | > |k_1\, v|$.
So, hereafter, we focus on the high frequency behavior
of the Poynting vector and we simply replace $\omega_-$ with $\omega$.

Since the Poynting vector $\Exp{T^{0i}(\omega,\bmx)}$
is defined by~\cite{GubserPufu07} as 
\be 
%S_i=-<T_{0i}>=\int \frac{d^3\bmk}{(2\pi)^3}\,\frac{k_i\omega}{k^2}
%Q_S e^{i\bmk\cdot \bmx}=-i\frac{\p}{\p x^i}
%\left[\int \frac{d^3\bmk}{(2\pi)^3}\frac{\omega}{k^2}
%Q_S e^{i\bmk\cdot \bmx}  \right], 
\Exp{T^{0i}(\omega,\bmx)}
=\int \frac{d^3k}{(2\pi)^3}~\frac{k_i\omega}{k^2}~Q\,
e^{i\bmk \cdot \bmx}
=-i\, \frac{\p}{\p x^i}~\left[ \int \frac{d^3k}{(2\pi)^3}~
\frac{\omega}{k^2}~Q\, e^{i\bmk\cdot \bmx}~\right]~, 
\ee
%the total flux $P$ from the two-dimensional spacelike surface $\p V$ enclosing the quark is 
%given by 
the total instantaneous power radiated by the quark
is given by 
\be 
P(t)=\oint_{\p V} \Exp{T^{0i}(t,\bmx)}\, n_i~, 
\ee
where $n_i$ is a unit normal vector
to two-dimensional spacelike surface ${\p V}$ enclosing the quark.
The first order terms of $v$ in Eq.~(\ref{qs})
does not contribute to the total instantaneous power $P$,
since $\Exp{T^{0i}}\, n_i \leftrightarrow - \Exp{T^{0i}}\, n_i$
for $x_1\leftrightarrow -x_1$ for the first order terms.
%Therefore, we only calculate the second order terms of $v$ and observe the angular dependence of 
%$\Exp{T^{0i}}\, n_i$.
%%Therefore, it is interesting to study the contribution from the second order terms of $v$
%%to the Poynting vector.
Therefore, we only calculate the second order terms of $v$.

After a little bit tedious calculation~
(in detail, see Appendix \ref{app:Poynting}),
one gets the Poynting vector $\Exp{T^{0i}}$ as
\be 
\label{qs2}
\Exp{T^{0i}(\omega,\bmx)}
%&=&-i\frac{\p}{\p x^i}\left[\int \frac{d^3\bmk}{(2\pi)^3}\frac{\omega}{k^2}
%Q_S e^{i\bmk\cdot \bmx}  \right] \nonumber \\
&=& ( \text{terms up to $v$} )
\nonumber \\
& &-\frac{iL^2v^2}{4\alpha'}\, \frac{\p}{\p x^i}
\left(\frac{I_0(R)}{3}-\frac{1}{3\omega^2}\frac{\p^2}{\p x_1^2}I_0(R)
+\frac{1}{2}\frac{\p^2}{\p x_1^2}I_2(R)
-\omega^2I_2(R)-\frac{3\omega^2}{2}\frac{\p^2}{\p x_1^2}I_4(R)\right)
\nonumber \\
&&+O(v^3)~,
\ee
where $R=|\bmx|$ and 
$I_0$, $I_1$, $I_2$, and $I_4$ are functions of $R$ and $\omega$
defined by
\begin{align}
  & I_0(R)
  = \frac{\omega}{4\pi R}\, \times \begin{cases}
      H_{1}^{(1)}(|\omega| R) & \text{~~for~~} \omega>0 \\
      H_{1}^{(2)}(|\omega| R) & \text{~~for~~} \omega<0 \end{cases}~,
\nonumber \\
  & I_1(R)
  = \frac{i}{4\pi R}\, \times \begin{cases}
      - H_{0}^{(1)}(|\omega| R) & \text{~~for~~} \omega>0 \\
        H_{0}^{(2)}(|\omega| R) & \text{~~for~~} \omega<0 \end{cases}~,
%\nonumber \\
%  & I_2(R)
%  = \frac{1}{4\pi  R|\omega|}\, \times \begin{cases}
%      \displaystyle{ \int^{|\omega|R}_0 dz~H_{0}^{(1)}(z) }
%    & \text{~~for~~} \omega>0 \\
%      \displaystyle{ - \int^{|\omega|R}_0 dz~H_{0}^{(2)}(z) }
%    & \text{~~for~~} \omega<0 \end{cases}~,
\nonumber \\
  & I_2(R)
  = \frac{1}{4\pi  R|\omega|}\, \int^{|\omega|R}_0 dz~\times
    \begin{cases}
      ~H_{0}^{(1)}(z) & \text{~~for~~} \omega>0 \\
      ~\left( - H_{0}^{(2)}(z) \right)
    & \text{~~for~~} \omega<0 \end{cases}~,
\nonumber \\
  & I_4(R)
  = - \frac{1}{R}\, \left[~
    \int^R \left( \int^{\hat{r}} \tilde{r} I_2(\tilde{r}) d\tilde{r}
           \right) d\hat{r} + C_1(\omega) R + C_2(\omega)~\right]~,
\end{align}
%\be
%I_0(R)&=&\left\{
%    \begin{array}{@{\,}ll} 
%         \frac{\omega}{4\pi r}H_{1}^{(1)}(|\omega| R)   
% & \mbox{for} \quad \omega>0 \\
%          \frac{\omega}{4\pi r}H_{1}^{(2)}(|\omega| R) 
%% &  \mbox{for} \quad \omega<0 \\         
%    \end{array} \right.,   \nonumber \\      
%I_1(R)&=&\left\{
%    \begin{array}{@{\,}ll} 
%         -\frac{i}{4\pi r}H_{0}^{(1)}(|\omega| R)   
% & \mbox{for} \quad \omega>0 \\
%          \frac{i}{4\pi r}H_{0}^{(2)}(|\omega| R) 
% &  \mbox{for} \quad \omega<0 \\         
%    \end{array} \right., \nonumber \\
%I_2(R)&=&\left\{
%    \begin{array}{@{\,}ll} 
%         \frac{1}{4\pi  R|\omega|}\int^{|\omega|r}_0H_{0}^{(1)}(z)dz   
% & \mbox{for} \quad \omega>0 \\
%          -\frac{1}{4\pi R|\omega|}\int^{|\omega|r}_0H_{0}^{(2)}(z)dz 
%% &  \mbox{for} \quad \omega<0 \\         
%    \end{array} \right., \nonumber \\ 
%I_4(R)&=&-\frac{1}{R}
%\left[\int^R\left(\int^{\hat{r}} \tilde{r}I_2(\tilde{r})d\tilde{r}\right)d\hat{r}
%+C_1(\omega) R+C_2(\omega)\right]
%\ee
with two arbitrary functions $C_1$ and $C_2$ of $\omega$.

In the asymptotic region $R \to \infty$,
the Hankel functions $H_0^{(n)}(|\omega| R)$ behaves as
\begin{align}
  & H_0^{(1)}(|\omega| R)
  \simeq \sqrt{\frac{2}{\pi}}~
  \frac{e^{i|\omega| R}}{\sqrt{|\omega| R}}~,
& & H_0^{(2)}(|\omega| R)
  \simeq \sqrt{\frac{2}{\pi}}~
  \frac{e^{-i|\omega| R}}{\sqrt{|\omega| R}}~,
\nonumber \\
  & H_1^{(1)}(|\omega| R)
  \simeq - i \sqrt{\frac{2}{\pi}}~
    \frac{e^{i|\omega| R}}{\sqrt{|\omega| R}}~,
& & H_1^{(2)}(|\omega| R)
  \simeq i \sqrt{\frac{2}{\pi}}~
    \frac{e^{-i|\omega| R}}{\sqrt{|\omega| R}}~.
\end{align}
So, in the frequency $\omega$-space,
the leading order of $\Exp{T^{0i}(\omega, \bmx)}$ becomes
%\be 
%\label{poynting3}
%\Exp{T^{0i}(\omega, R)} \simeq\left\{
%    \begin{array}{@{\,}ll} 
%  %   +\frac{iL^2v^2\omega^2 x^i}{24\pi\alpha' r^2}\left(\frac{2x^2}{r^2}-1\right)H_0^{(1)}(|\omega|r)
%  -\frac{iL^2v^2{|\omega|}^{3/2} x^i}{12\sqrt{2}\pi^{3/2}\alpha' R^{5/2}}\left(\frac{2x_1^2}{R^2}-1\right)
%   e^{i|\omega| R}=O\left(\frac{\omega^{3/2}}{R^{3/2}}\right),  & \,\,\omega>0 \\
%   % -\frac{iL^2v^2\omega^2 x^i}{24\pi\alpha' r^2}\left(\frac{2x^2}{r^2}-1\right)H_0^{(2)}(|\omega|r), 
%    -\frac{iL^2v^2{|\omega|}^{3/2} x^i}{12\sqrt{2}\pi^{3/2}\alpha' R^{5/2}}\left(\frac{2x_1^2}{R^2}-1\right)
%   e^{-i|\omega| R}=O\left(\frac{\omega^{3/2}}{R^{3/2}}\right),   & \,\,\omega<0.  
%       \end{array}
%        \right.  
%\ee
\be 
\label{poynting3}
   \Exp{T^{0i}(\omega, R)}
  \simeq - \frac{iL^2v^2{\omega} x^i}
                {12\pi\alpha' R^2}
   \left(\frac{2x_1^2}{R^2}-1\right) \sqrt{\frac{|\omega|}{2\pi R}}e^{i \omega R}
 = O\left( \frac{|\omega|^{3/2}}{R^{3/2}} \right)~.
\ee
The factor $e^{i \omega R}$ shows that
the energy flux is an outgoing wave, as we expect.
Since this factor oscillates heavily
with respect to $\omega$ for large $R$, we might expect 
that the slow falloff with $R$ in Eq.~(\ref{poynting3}) 
disappears in position space except the neighborhood of 
future light cone $t\sim R$. In fact, the first term in Eq.~(\ref{qs2}), for example, 
becomes in position space 
\be 
\left| \int^\infty_{-\infty} d\omega~e^{-i\omega t}\,
\frac{\p I_0(R )}{\p x^i} \right|
\sim \frac{|x^i|\theta(t-R)}{(t^2-R^2)^{5/2}}\le
\frac{R\,\theta(t-R)}{(t^2-R^2)^{5/2}}\le
\frac{\theta(t-R)}{t^4 (1-(R/t)^2)^{5/2}}~, \nonumber
\ee    
where the step function inevitably appears from the causality condition. 
It is easily seen that in any timelike directions $(t>R)$ the term rapidly decays 
as $t^{-4}$. One can also see that the other terms in Eq.~(\ref{qs2}) decays similarly. 
So, the slow falloff with $R$ in Eq.~(\ref{poynting3}) disappears in the real 
space.~\footnote{Since the expansion of Eq.~(\ref{suddenly-coe}) by $v$
breaks down near the zero mode, we cannot strictly derive
the energy flux in position space. From the dimensional analysis, however,
we expect the power law decay in time.} The strong divergence on the light cone $t\sim R$ 
may be due to the simple model in which the string suddenly stops. 

In terms of 't Hooft coupling $g_{\text{YM}}^2N$, Fourier transform of 
the total instantaneous power $P(\omega)$ is obtained from Eq.(\ref{poynting3}) as
\be
\label{poynting}
P(\omega) = \oint_{\partial V} \Exp{T^{0i}(\omega, R)}\, n_i
  \simeq \frac{iv^2\omega}{9}
  \sqrt{|\omega|\frac{g_{\text{YM}}^2N}{2\pi}}
   R^{1/2}\, e^{i \omega R}. 
   %\int^{\pi}_{0} d\Theta~\sin\Theta\, \cos2\Theta~,
%
%\left\{
%    \begin{array}{@{\,}ll} 
%-\frac{iv^2|\omega|^{3/2}\sqrt{g_{YM}^2N}}{12\sqrt{2}\pi^{3/2}}\frac{e^{i|\omega|r}}{R^{3/2}} \cos2\Theta,
%& \,\,\omega>0 \\
%-\frac{iv^2|\omega|^{3/2}\sqrt{g_{YM}^2N}}{12\sqrt{2}\pi^{3/2}}\frac{e^{-i|\omega|r}}{r^{3/2}}\cos2\Theta,
%& \,\,\omega<0
%   \end{array}
%        \right. 
\ee
%where $\Theta$ is the angle between $x_1$-axis and $\bmx$.

It is interesting to compare Eq.(\ref{poynting})
%the angular distribution of the Poynting flux $\Exp{T^{0i}}\, n_i$
with the known result in the case of electromagnetism.
%c field in the same frequency $\omega$.  
In the electromagnetic case,
we consider the bremsstrahlung radiation sourced by an electron
with initial velocity $v$ which rapidly stops
in the interval $\Delta t$.
In the low velocity limit $v\to 0$, Fourier transform of 
the total instantaneous power radiated is given by
\be
\label{poynting1}
P(\omega) = \oint \Exp{T^{0i}(\omega,r)}\, n_i
 \simeq - \frac{2e^2v^2}{3\Delta t}~e^{i \omega R}.  
%\int^\pi_0 d\Theta~\sin\Theta\, (1-\cos2\Theta)~,
\ee
%where $e$ and $c$ represent the charge of the electron
%and the speed of light, respectively.
Eq.~(\ref{poynting}) does not include the time interval $\Delta t$
since the quark suddenly stops with almost zero time interval.
So, we must compare Eq.~(\ref{poynting}) with Eq.~(\ref{poynting1})
in the range $|k_1 v| \ll |\omega | \ll 1/\Delta t$.
We find that both the coupling dependence
and the frequency dependence are quite different
between the strongly coupled gauge theories and electromagnetism.
% at least in the region of $|k_1 v| \ll |\omega | \ll 1/\Delta t$.
%%%%%%%%%%%%%%%%%%%%%%%%%%%%%%%%%%%
\section{conclusion and discussions}
\label{sec:conclusion}
%%%%%%%%%%%%%%%%%%%%%%%%%%%%%%%%%%%
We have examined radiation by an accelerated quark
in ${\cal N}=4$ SYM theory
with large 't Hooft coupling via gauge/gravity duality.
In the oscillating quark with frequency $\Omega$,
we have shown that the time averaged energy density
far away from the quark correctly falls off as $\Omega^4/R^2$.  
%This falloff demonstrates the existence of SYM radiation   
%the variation of the time averaged energy density
%with $\Omega$ and distance $R$ is given by $\Omega^4/R^2$.
% asymptotically.
The $\Omega$ dependence also agrees with the result in electromagnetism.

%Since the strength of the acceleration of the quark $\Dot{v}$ is roughly $\Omega^2$,  
%the energy density can be expressed by $\Dot{v}^2/R^2$. This is the expected result 
%by analogy with electromagnetic field.
Interestingly, the angular distribution
of the time averaged energy density is isotropic. Since ${\cal N}=4$ SYM theory 
is in deconfined phase, gluons can be emitted. 
Although the energy density is composed of not only gluons,
but also other fields, it is natural to consider that the dominant ingredient of 
the energy density is gluon. In the weak coupling limit, the gluon radiation 
should become dipole radiation analogously with electromagnetism. 
So, one may expect that dipole radiation occurs even in ${\cal N}=4$ SYM theory. 
%since emitted particles are in color singlet states. however, gluons can be emitted.  
%Although the energy density is composed of gluons
%and many kind of scalar fields, it is natural to consider that
%the dominant ingredient of the energy density is gluon.
%So, by analogy with electromagnetism,
%the radiation should approach dipole radiation
%in the weak coupling limit of ${\cal N}=4$ SYM.
%This suggests that the isotropic energy density
%is induced by the effect of strong coupling in the following:
%In the strongly coupled gauge theory a gluon induces
%%production of gluons around it by the self-interaction,
%and as a result, the energy density
%is asymptotically averaged over all the angles.
%The reason why isotropic radiation 
%occurs in the strong coupling limit may be explained as follows:  
%while the radiation would be isotopic in the strong coupling limit.
%Even for the gluon radiation, we can expect that
%the radiation is isotropic in the strong couplig limit as follows:
In the strongly coupled gauge theory, however, a gluon emitted by the source quark induces 
production of gluons around it by the self-interaction. As a result, we may expect that 
the energy density is asymptotically averaged over all the angles by the gluons.    
%Suppose that the primary gluons emitted by an oscillating quark propagate  
%along the direction perpendicular to the trajectory of the quark. 
%Owing to the self-interaction of gluons,
%a primary gluon can emit the secondary gluons which propagate along the 
%direction perpendicular to the trajectory of the primary gluon. 
%so that the secondary gluons distribute along the direction of
%the trajectory of the source quark.
%Thus, the self-production process of emitted gluons
%make distribution of radiation isotropic in the strongly coupled gauge theory.}

%the gluon radiation could expand in every direction 
%by ``cascading decay'' process and, as a result, an isotropic energy density could be generated at 
%infinity. 
%isotropic radiation would be understood
%that scalar glueball radiation is dominant.
%However, $\mathcal{N}=4$ SYM theory has no confined phase, $\cdots$
%Additionaly, we note that the constituent of the energy density is not only 
%gluons but also scalar fields.
%Although the constituent of the energy density is not only gluons but also scalar fields,
%the isotropic distribution is characteristic in the strongly coupled gauge field.      

In the toy model of a scattered quark,
we derived the total instantaneous power sourced by a quark
which moves at constant velocity $v$,
suddenly stops at the origin and then remains at rest
by calculating Poynting vector $\Exp{T^{0i}}$
via gauge/gravity duality.
In this case, we found that both the coupling dependence
and the frequency dependence are quite different
from the case of electromagnetism.
The Fourier transform of the total instantaneous power
in electromagnetism $P(\omega)$ is independent of the frequency $\omega$, 
while $P(\omega)$ in $\mathcal{N}=4$ SYM increases with $\omega$. 
%It implies that the variation in time of the total instantaneous power is heavily 

%The dominant contribution in the frequency is high frequency mode in the gauge theory,
%while every frequency mode is equally contributed in the flux in 
%electromagnetic field. 

Even though the static force in strongly coupled gauge theory
is well understood in terms of Wilson line probe,
the radiation problem is still largely open question.
In this paper we have investigated properties of radiation
at zero temperature as a first step toward the understanding of the 
properties at finite temperature. 
It is still an open question whether or not universality exists in the radiation 
properties, although we have seen it in the ratio of 
viscosity divided by entropy density. If universality exists, we expect that the 
investigation at finite temperature may be connected to the RHIC data through jet 
quenching phenomenon. 
Even though there is no universality in the properties of radiation,    
%Since supersymmetry and conformal symmetry are irrelevant
%at finite temperature, ${\cal N}=4$ SYM and QCD
%just above the deconfinement temperature
%share much in common~\cite{ShuryakZahed1,ShuryakZahed2}.
we would obtain some insight into the understanding of the mechanism of 
energy dissipation by an external quark moving through the thermal plasma.

%%%%%%%%%%%%%%%%%%%%%%%%%%%%%%%%%%%%%%%%%%%%%%%%%%%%%%%%%%%%%%%%%%%%%%%%%%%%
%Motivated by the phenomenon of jet-quenching observed in RHIC experiments,
%a large number of studies~\cite{HKKKY,gubser06,herzog06,LiuRajWie06,CaTe06,Buchel06,CacGui06} 
%have been made in deriving energy dissipation rate of quark passing through the plasma.
%To know about the mechanism of jet-quenching mechanism,
%it is necessary to study the detail of radiation pattern.
%%little is known about the mechanism of energy dissipation.
%In the papers \cite{GubserYarom0709,GPY07064,GPY07060,GubserPufu07,FGMP06},
%the radiation pattern is studied for a quark moving at constant velocity.
%These researches correspond to \lq\lq gluon radiation mechanism\rq\rq which is one of jet-quenching scenarios.

%we should consider
%If energy loss results from gluons 
%radiating off sourced by a quark moving in a zigzag or being scattered by one of 
%constituents in thermal plasma. So, the basic question is: How much energy is the 
%quark radiating in the thermal plasma? How is the angular distribution of the energy 
%flux? We hope that the next detailed investigation about the radiation sourced by an 
%accelerated quark in thermal plasma is exploited to understand the mechanism of 
%the energy dissipation.     
%%%%%%%%%%%%%%%%%%%%%%%%%%%%%%%%%%%%%%%%%%%%%%%%%%%%%%%%%%%%%%%%%%%%%%%%%%%%

\begin{acknowledgments}
We would like to thank M. Natsuume for useful discussions. We are grateful to K. Matsui 
for comments. 
\end{acknowledgments}

%%%%%%%%%%%%%%%%%%%%%%%%%%%%%%%%%%%%%%%
\appendix
%%%%%%%%%%%%%%%%%%%%%%%%%%%%%%%%%%%%%%%
\section{Calculation of coefficient $Q$ for oscillating string}
\label{app:coeff_B}
%%%%%%%%%%%%%%%%%%%%%%%%%%%%%%%%%%%%%%%
Since the coefficient $Q$ in Eq.(\ref{eq:falloff})
is finite part of $B$ in Eq.(\ref{coefficient}),
we give calculation of $B$ in this appendix.

Let us define $j_{0C}$, $j_{0S}$, and $j_{0N}$ as
\be 
j_{0C}&:=&\epsilon^2\left(\frac{C_1}{r^3}+\frac{C_2}{r^5}
+\frac{C_3}{r^7}\right)\cos(2\Omega z)~,
\nonumber \\  
j_{0S}&:=&\epsilon^2\left(\frac{S_1}{r^4}+\frac{S_2}{r^6} \right)
\sin(2\Omega z)~,
\nonumber \\
j_{0N}&:=&\epsilon^2\left(\frac{N_1}{r^3}+\frac{N_2}{r^5}
+\frac{N_3}{r^7} \right)~.
\ee
The coefficient $B_S$ coming from the partial source $j_{0S}$
can be calculated as 
\be 
B_S(r)&=&\epsilon^2\int^r_0 dr_0~\left(\frac{S_1}{r_0}+\frac{S_2}{r_0^3}
\right)K_0(kL^2/r_0) \sin(2\Omega L^2/r_0)
\nonumber \\
&=&S_1\epsilon^2\sqrt{\pi L^2\Omega}
\int^\infty_{1/r}dz_0~z_0^{-1/2}K_0(kL^2z_0)J_{1/2}(2L^2\Omega z_0)
\nonumber \\
&&+S_2\epsilon^2\sqrt{\pi L^2\Omega}
\int^\infty_{1/r}dz_0~z_0^{3/2}K_0(kL^2z_0)J_{1/2}(2L^2\Omega z_0). 
\ee
So, in the large $r$ limit, we obtain the coefficient $B_S$ as 
\be
\nonumber \\
B_S&=&\frac{\pi}{2}S_1\epsilon^2
\ln\left(\frac{2\Omega}{k}+\sqrt{\frac{4\Omega^2}{k^2}+1}\right)
+\frac{\pi\Omega S_2\epsilon^2}{L^4\left(k^2+4\Omega^2 \right)^{3/2}}~
\ee
with the help of Appendix C.

The coefficient $B_N$ coming from the partial source $j_{0N}$
can be calculated as
\be 
B_N(r)&=&\epsilon^2\int^r_0 dr_0~
\left(N_1+\frac{N_2}{r_0^2}+\frac{N_3}{r_0^4} \right) K_0(kL^2/r_0)
\nonumber \\
&=&\epsilon^2\int^\infty_{kL^2/r} dy~
\left(\frac{kL^2N_1}{y^2}+\frac{N_2}{kL^2}+\frac{N_3y^2}{(kL^2)^3}
\right)K_0(y)
\nonumber \\
&=&-\epsilon^2N_1kL^2\left(\left[\frac{K_0(y)}{y} \right]^\infty_{kL^2/r}
+\int^\infty_{kL^2/r} dy~\frac{K_1(y)}{y} \right)
\nonumber \\
&&+\frac{\epsilon^2N_2}{kL^2} \int^\infty_{kL^2/r} dy~K_0(y)
+\frac{\epsilon^2N_3}{(kL^2)^3}
\int^\infty_{kL^2/r} dy~y^2 K_0(y)
\nonumber \\
%&=&-\epsilon^2N_1kL^2\left(\left[\frac{K_0(y)}{y}  \right]^\infty_{kL^2/r}
%+\frac{1}{2}\int^\infty_{kL^2/r}K_2(y)dy  
%-\frac{1}{2}\int^\infty_{kL^2/r}K_0(y)dy\right) \nonumber \\
%&&+\frac{\epsilon^2N_2}{kL^2}\int^\infty_{kL^2/r}K_0(y)dy+\frac{\epsilon^2N_3}{(kL^2)^3}
%\int^\infty_{kL^2/r} y^2K_0(y)dy \nonumber \\
&=&-\epsilon^2N_1kL^2\left(\left[\frac{K_0(y)}{y}  \right]^\infty_{kL^2/r}
-\left[K_1(y)  \right]^\infty_{kL^2/r}\right)
\nonumber \\
&&+\epsilon^2\left(N_1kL^2+\frac{N_2}{kL^2}\right)
\int^\infty_{kL^2/r} dy~K_0(y) + \frac{\epsilon^2N_3}{(kL^2)^3}
\int^\infty_{kL^2/r} dy~y^2 K_0(y)~,
\ee
where we used the formulas 
\begin{align}
  & K_0(y)-K_2(y)=-\frac{2}{y}K_1(y)~,
& & K_0(y)+K_2(y)=-2K'_1(y)~,
\end{align}
in the third and fourth lines, respectively.
In the large $r$ limit~($r\to \infty$), we obtain the coefficient $B_N$ as
\be
B_N=\frac{\pi}{2}\epsilon^2
\left(kL^2N_1+\frac{N_2}{kL^2}+\frac{N_3}{(kL^2)^3}\right) +O(r\ln r)~,
\ee
with the help of formulas
$\int^\infty_0K_0(y)dy=\int^\infty_0y^2K_0(y)dy=\pi/2$.

The coefficient $B_C$ coming from the partial source $j_{0C}$
can be calculated in the large $r$ limit as
\be 
B_{C}(r)&=&\epsilon^2\int^r_0 dr_0~
\left(C_1+\frac{C_2}{r_0^2}+\frac{C_3}{r_0^4} \right)
K_0(kL^2/r_0) \cos(2L^2\Omega/r_0)
\nonumber \\
&=&\frac{\epsilon^2kL}{2}\sqrt{\frac{\pi}{\Omega}}
\int^\infty_{1/r} dz_0~(C_2z_0^{1/2}+C_3z_0^{5/2})
K_1(kL^2 z_0)J_{1/2}(2L^2\Omega z_0)
\nonumber \\
&-&\frac{\epsilon^2C_3}{L}\sqrt{\frac{\pi}{\Omega}}
\int^\infty_{1/r} dz_0~z_0^{3/2} K_0(kL^2 z_0) J_{1/2}(2L^2\Omega z_0)
+\epsilon^2C_1L^2\int^\infty_{L^2/r}dz_0~
\frac{K_0(kz_0)}{z_0^2}\cos(2\Omega z_0)
\nonumber \\
&\simeq&\frac{\pi\epsilon^2 C_2}{2L^2\sqrt{k^2+4\Omega^2}}
+\frac{\pi\epsilon^2C_3(k^2-8\Omega^2)}
{2L^6\left(k^2+4\Omega^2\right)^{5/2}}
+\epsilon^2C_1L^2
\int^\infty_{L^2/r}dz_0~\frac{K_0(kz_0)}{z_0^2}\cos(2\Omega z_0)~,
\ee
with the help of Appendix C. 
The third term in the last line includes divergent terms.
So, using Eq.~(\ref{integral-K}), we carefully calculate as
\be 
P(r)&:=&\epsilon^2C_1L^2\int^\infty_{L^2/r}dz_0~
\frac{K_0(kz_0)}{z_0^2}\cos(2\Omega z_0)
\nonumber \\
&=&\epsilon^2C_1r\cos b\int^\infty_0 dt~\frac{\cos(at)}{\sqrt{t^2+1}}
+\frac{\epsilon^2C_1r}{2}\int^\infty_0 \frac{dt}{\sqrt{t^2+1}}
\{(at+b)\mbox{si}(at+b)+(at-b)\mbox{si}(at-b)\}
\nonumber \\
&\simeq&\epsilon^2C_1 r\cos bK_0(a)
-\epsilon^2C_1r\int^\infty_0 \frac{dt}{\sqrt{t^2+1}}
\left\{at\,\mbox{si}(at)-\frac{b^2}{2}\cos(at)
-\frac{b^2}{2t}\sin(at)\right\}
\nonumber \\
&=&\epsilon^2C_1r\left(K_0(a)\cos b +a\int^\infty_a dx~\frac{K_1(x)}{x}
+\frac{b^2}{2}K_0(a)+\frac{b^2}{2a}\int^a_0 dx~K_0(x) \right)~,
\label{eq:P_in_App_A}
\ee
where $a=kL^2/r$ and $b=2\Omega L^2/r$
and the function $\mbox{si}(x)$ is defined by
\begin{align}
  & \mbox{si}(x) := - \int^\infty_x dt~\frac{\sin t}{t}~.
\end{align}
In the third line,
we used the Taylor expansion of $\mbox{si}(x)$ around $x=at$
for large $r$.
In the asymptotic region, $r\to \infty$, we obtain
\be 
P\sim O(r\ln r). 
%\epsilon^2C_1\left(\cos b K_0(a) r+1+O\left(\frac{1}{r^2}\right)\right)~.
\ee  
This logarithmic divergence does not appear in Eq.~(\ref{asymptotics}) because the coefficient 
$A$ in Eq.~(\ref{coefficient}) also includes the same divergence with an opposite sign. 
So, $P$ does not contribute to $Q(\omega, {\bf k})$ in Eq.~(\ref{asymptotics}). 
%%%%%%%%%%%%%%%%%%%%%%%%%%%%%%%%%%%%%%%%%%
\section{Calculation of $\Exp{T^{0i}}$ for a scattered string}
\label{app:Poynting}
%%%%%%%%%%%%%%%%%%%%%%%%%%%%%%%%%%%%%%%%%%
The calculation of $\Exp{T^{0i}}$ can be easily
computed by differentiating appropriately modified versions
of the following Fourier transforms of $1/q$ and $1/kq$:
\be 
I_0(R)&:=&\frac{1}{(2\pi)^3}\int d^3k~\frac{e^{i\bmk\cdot \bmx}}{q}
\nonumber \\
&=&\frac{1}{(2\pi)^2iR}
\int^\infty_{-\infty} dk~\frac{ke^{ikR}}{q}
\nonumber \\
&=&\frac{1}{(2\pi)^2iR} \left[~\int^{|\omega|}_{-|\omega|} dk~
\frac{ke^{ikR}\mbox{sgn}(\omega)}{\sqrt{\omega^2-k^2}}
+\int^\infty_{-\infty} dk~
\frac{\theta(|k|-|\omega|)ke^{ikR}}{i\sqrt{k^2-\omega^2}}~\right]
\nonumber \\
&=&\frac{1}{(2\pi)^2iR}\left[~\pi i\omega J_1(|\omega| R)
-\frac{\p}{\p r}\int^\infty_{-\infty} dk~
\frac{\theta(|k|-|\omega|)e^{ikR}}{\sqrt{k^2-\omega^2}}~\right]
\nonumber \\
&=&\frac{1}{(2\pi)^2iR}\left[~\pi i\omega J_1(|\omega|R)
+\pi\frac{\p}{\p R}N_0(|\omega| R)~\right]
\nonumber \\
&=& \frac{\omega}{4\pi R}\, \times \begin{cases}
    H_{1}^{(1)}(|\omega| R) & \text{~~for~~} \omega>0 \\
    H_{1}^{(2)}(|\omega| R) & \text{~~for~~} \omega<0 \end{cases}~,
\nonumber \\
I_1(R)&:=&\frac{1}{(2\pi)^3} \int d^3k~\frac{e^{i\bmk\cdot \bmx}}{q k}
\nonumber \\
%&=&\frac{1}{(2\pi)^2ir}\int^\infty_{-\infty}\frac{ke^{ikr}}{q}dk \nonumber \\
&=&\frac{1}{(2\pi)^2iR} \left[~\int^{|\omega|}_{-|\omega|} dk~
\frac{e^{ikR}\mbox{sgn}(\omega)} {\sqrt{\omega^2-k^2}}
+ \int^\infty_{-\infty} dk~
\frac{\theta(|k|-|\omega|)e^{ikR}}{i\sqrt{k^2-\omega^2}}~\right]
\nonumber \\
&=&\frac{1}{(2\pi)^2iR}\left[~\pi \text{sgn}(\omega) J_0(|\omega| R)
+\pi i N_0(|\omega| R)~\right]
\nonumber \\
&=& \frac{i}{4\pi R}\, \times \begin{cases}
    - H_{0}^{(1)}(|\omega| R) & \text{~~for~~} \omega>0 \\
      H_{0}^{(2)}(|\omega| R) & \text{~~for~~} \omega<0 \end{cases}~,
\ee
where $R=|\bmx|$.
So, Fourier transformation of $1/k^2q$ is easily calculated as
\be
I_2(R)&:=&\frac{1}{(2\pi)^3}
\int d^3k~\frac{e^{i\bmk\cdot \bmx}}{k^2q}
\nonumber \\
&=&\frac{1}{2\pi^2R} \int^\infty_0 dk~\frac{\sin(kR)}{k q}
\nonumber \\
&=&\frac{1}{2\pi^2R}\int^R_0 d\tilde{r}~
\left[~\int^\infty_0 dk~\frac{\cos(k\tilde{r})}{q}~\right]
\nonumber \\
&=&\frac{i}{R}\int^R_0 d\tilde{r}~\tilde{r}I_1(\tilde{r})
%\nonumber \\
%&=& \frac{1}{4\pi R|\omega|}\, \times \begin{cases}
%   \displaystyle{ \int^{|\omega| R}_0 dz~H_{0}^{(1)}(z) }
%     & \text{~~for~~} \omega>0 \\
%   \displaystyle{ - \int^{|\omega| R}_0 dz~H_{0}^{(2)}(z) }
%     & \text{~~for~~} \omega<0 \end{cases}~.
\nonumber \\
&=& \frac{1}{4\pi R|\omega|}\, \int^{|\omega| R}_0 dz~\times
\begin{cases}
  ~H_{0}^{(1)}(z) & \text{~~for~~} \omega>0 \\
  ~\left( - H_{0}^{(2)}(z) \right) & \text{~~for~~} \omega<0
\end{cases}~.
\ee
Similarly, from the fact that
\be
\frac{\p^2}{\p R^2}\left[~\frac{R}{(2\pi)^3}
\int d^3k~\frac{e^{i\bmk\cdot \bmx}}{k^4q}~\right]
=- R I_2(R)~,
\ee
Fourier transformation of $1/k^4q$ can be represented
by two integration constants $C_1(\omega)$ and $C_2(\omega)$ as
\be 
I_4(R)&:=&\frac{1}{(2\pi)^3} \int d^3k~\frac{e^{i\bmk\cdot \bmx}}{k^4q}
\nonumber \\
&=&-\frac{1}{R} \left[~\int^R d\hat{r}~
\left( \int^{\hat{r}} d\tilde{r}~\Tilde{r}\, I_2(\tilde{r}) \right)
+ C_1(\omega)r + C_2(\omega)~\right]~.
\ee

%%%%%%%%%%%%%%%%%%%%%%%%%%%%%%%%%
\section{Formulas}
\label{app:formulas}
%%%%%%%%%%%%%%%%%%%%%%%%%%%%%%%%%
%For any parameters $\lambda$, $\mu$, $\nu$, $a$, and $b$ satisfying
%\be 
%\mbox{Re}(\lambda\pm\mu+\nu)>-1, \qquad |\mbox{Im}(b)|<\mbox{Re}(a), \nonumber 
%\ee 
%\be 
%\int^\infty_0x^\lambda K_\mu(ax)J_\nu(bx)dx&=&
%\frac{2^{\lambda-1}(ib/a)^\nu}{a^{\lambda+1}\Gamma(\nu+1)}
%\Gamma\left(\frac{\lambda+\mu+\nu+1}{2}\right)
%\Gamma\left(\frac{\lambda-\mu+\nu+1}{2}\right)\times \nonumber \\
%&&F((\lambda+\mu+\nu+1)/2,(\lambda-\mu+\nu+1)/2,\nu+1;-b^2/a^2), 
%\ee
%where $F$ is the hypergeometric function. In the specific case, one obtains 
For any real positive parameter $a$ and $b$, one obtains \cite{Watson}
\begin{align}
\label{integral-J}
& \int^\infty_0 dx~x^{-1/2}\, K_0(ax)\, J_{1/2}(bx)
= \sqrt{\frac{\pi}{2b}}~\sinh^{-1}\left(\frac{b}{a}\right)
= \sqrt{\frac{\pi}{2b}}~\ln\left( \frac{b}{a}
  + \sqrt{\frac{b^2}{a^2}+1} \right)~,
\nonumber \\
& \int^\infty_0 dx~x^{3/2}\, K_0(ax)\, J_{1/2}(bx)
= \sqrt{\frac{\pi b}{2}}~
%  \frac{1}{a^3(1+b^2/a^2)^{3/2}}~,
  \frac{1}{(a^2+b^2)^{3/2}}~.
\end{align}

It is easy to get the following equations
by differentiating Eq.(\ref{integral-J}) with respect to $a$,
\begin{align}
\label{integral-JII}
& \int^\infty_0 dx~x^{1/2}K_1(ax)J_{1/2}(bx)
= \sqrt{\frac{\pi b}{2}}~\frac{1}{a(a^2+b^2)^{1/2}}~,
\nonumber \\
& \int^\infty_0 dx~x^{5/2}\, K_1(ax)\, J_{1/2}(bx)
= \sqrt{\frac{\pi b}{2}}~\frac{3a}{(a^2+b^2)^{5/2}}~.
\end{align}

In Eq.(\ref{eq:P_in_App_A}),
we use the integral representation of $K_0(x)$
\be 
\label{integral-K}
K_0(x)=\int^\infty_0 dt~\frac{\cos(xt)}{\sqrt{t^2+1}}~.
\ee

%%%%%%%%%%%%%%%%%%%%%%%%% references %%%%%%%%%%%%%%%%%%%%%%%%
%%%%%%%%%%%%%%%%%%%%%%%%%%%%%%%%%%%%%%%%%%%%%%%%%%%%%%%%%%%%%

\end{document}